\documentclass[prl,twocolumn,aps,showpacs,superscriptaddress]{revtex4-1}

\usepackage{graphicx}
\usepackage{bm}
\usepackage{amsmath}
\usepackage{amssymb}
\usepackage{textcomp}
\usepackage[utf8]{inputenc}
\usepackage[USenglish]{babel}
\usepackage{microtype}
\usepackage{color}
\usepackage[bookmarks=false]{hyperref}
\usepackage[all]{hypcap}
\newcommand{\figref}[1]{Fig.~\ref{#1}}
\newcommand{\intextcite}[1]{Ref.~\citenum{#1}}

\newcommand{\unit}[1]{\,\mathrm{#1}}
\newcommand{\rangstrom}{\,\mbox{\AA}^{-1}}

\newcommand{\tcdw}{T_{\mbox{\scriptsize CDW}}}

\newcommand{\omp}{\omega_p}

\newcommand{\twoh}{$2$\emph{H}}
\newcommand{\htase}{\twoh-TaSe$_2$}
\newcommand{\htas}{\twoh-TaS$_2$}
\newcommand{\hnbse}{\twoh-NbSe$_2$}
\newcommand{\hnbs}{\twoh-NbS$_2$}

\begin{document}

\title{The Effect of Charge Order on the Plasmon Dispersion in Transition-Metal Dichalcogenides}

\author{Jasper van Wezel}
\affiliation{Materials Science Division, Argonne National Laboratory, Argonne, IL 60439, USA}

\author{Roman Schuster}
\affiliation{IFW Dresden, P.O. Box 270116, D-01171 Dresden, Germany}
\affiliation{Department of Physics, University of Fribourg, CH-1700 Fribourg, Switzerland}

\author{Andreas König}
\author{Martin Knupfer}
\author{Jeroen van den  Brink}
\affiliation{IFW Dresden, P.O. Box 270116, D-01171 Dresden, Germany}

\author{Helmuth Berger}
\affiliation{Institut de Physique de la Matière Complexe, EPFL, CH-1015 Lausanne, Switzerland}

\author{Bernd Büchner}
\affiliation{IFW Dresden, P.O. Box 270116, D-01171 Dresden, Germany}

~\\

\begin{abstract}
We investigate the dispersion of the charge carrier plasmon in the three prototypical charge-density wave bearing transition-metal dichalcogenides \htase, \htas\ and \hnbse\ employing electron energy-loss spectroscopy. For all three compounds the plasmon dispersion is found to be negative for small momentum transfers. This is in contrast to the generic behavior observed in simple metals as well as the related system \hnbs, which does not exhibit charge order. We present a semiclassical Ginzburg-Landau model which accounts for these observations, and argue that the vicinity to a charge ordered state is thus reflected in the properties of the collective excitations.
\end{abstract}


\maketitle

{\it Introduction.}--Charge-density waves are well known to affect many of the single-particle properties of the materials in which they occur \cite{Gruner}. The experimental consequences include the appearance of superstructure reflections in elastic X-ray scattering, the softening of acoustic phonons and the appearance of an excitation gap in the single-particle spectrum \cite{Gruner,Kohn_Phys.Rev.Lett._1959}. Much less, however, is known about the possible influence of charge-density waves (CDW) on collective excitations. Since the generic CDW instability is accompanied by a strong enhancement of the electronic susceptibility, it is natural to expect that experimental probes which are sensitive to the susceptibility (and thus probe collective excitations), should reveal characteristic features of the CDW as well. It is the main aim of this Letter to show that this is indeed the case.

Many of the layered transition-metal dichalcogenides (TMDC) are considered to be prototypical CDW materials \cite{Leininger_Phys.Rev.B_2011,Moncton_Phys.Rev.B_1977,Borisenko_Phys.Rev.Lett._2008,JvW:PRB81,JvW:PRB83}. Because they are metallic, the conduction bands of the charge ordered TMDC host collective density oscillations, or plasmons, in addition to the single-particle states \cite{Pines_Rev.Mod.Phys._1956}. The typical energy required to excite such collective modes is given by the plasma frequency $\omp=\sqrt{e^2 n / \epsilon m}$, which scales with the density $n$ of the conduction electrons and easily reaches several $\unit{eV}$ for ordinary metals \cite{Schnatterly1979}. One may therefore be tempted to conclude that fluctuations of the CDW order, which occur on the $\unit{meV}$ scale, could hardly play a role for the plasmon dynamics, contrary to the anticipation mentioned above.

In the following we resolve this paradox and establish a connection between the plasmon dispersion and the presence of CDW order in the specific TMDC systems \htase, \htas\ and \hnbse\ (with $\tcdw \sim 120 \unit{K}$, $77 \unit{K}$ and $33 \unit{K}$ respectively). To this end we apply inelastic electron scattering  \cite{Schnatterly1979,Fink_J.ElectronSpectrosc.Relat.Phenom._2001} (often termed electron energy-loss spectroscopy or EELS) to study the interplay of (fluctuating) charge order with the plasmon mode of the charge carriers. We show that in all three compounds the plasmon has a negative dispersion at low values of the momentum transfer, leading to a minimum in the dispersion close to the CDW ordering wave vectors. This is in contrast to the generic description of a metal in a single-band model with a spherical Fermi surface, where a low-momentum expansion of the Lindhard function gives rise to a strictly positive plasmon dispersion \cite{Pines_1963}. 

A negative plasmon dispersion has been observed before in the heavy alkali metals \cite{Felde_Europhys.Lett._1987,Felde_Phys.Rev.B_1989}, where it has been attributed to the presence of interband as well as intraband transitions \cite{Aryasetiawan_Phys.Rev.Lett._1994,Paasch_Ukr.Fiz.Zhurn._1999}. In a Wigner crystal, where the plasmon mode is identical to an optical phonon mode, the plasmon dispersion is even necessarily negative \cite{Maksimov:2007ks}. For the TMDC considered here however, the electronic density is far larger than in a Wigner crystal. In addition, the closely related material \hnbs, which has a very similar band structure to the TMDC studied here but does not charge order, does have a strictly positive plasmon dispersion. Motivated by this observation we use a semiclassical approach based on a Ginzburg-Landau model for the charge ordered state to show that the presence of collective charge excitations associated with the vicinity of a CDW transition strongly affects the dynamics of the collective plasmon mode, and causes a dip in the plasmon dispersion close to the CDW ordering wavevector, as observed in the charge ordered TMDC.

{\it Experiment and Results.}--Single crystals were prepared according to the method described in \cite{Schuster_Phys.Rev.B_2009}. Thin films required for the transmission geometry were prepared from the single crystals with the help of an ultramicrotome or, whenever possible, by cleaving repeatedly with adhesive tape. The experiments were carried out in a purpose-built transmission EELS spectrometer \cite{Fink_Adv.Electron.ElectronPhys._1989}, equipped with a helium flow cryostat, and with the overall energy and momentum resolution set to $\Delta E=80\unit{meV}$ and $\Delta q=0.035\rangstrom$ respectively.
\begin{figure}[t]  
\centering
  \includegraphics[width=0.8 \columnwidth]{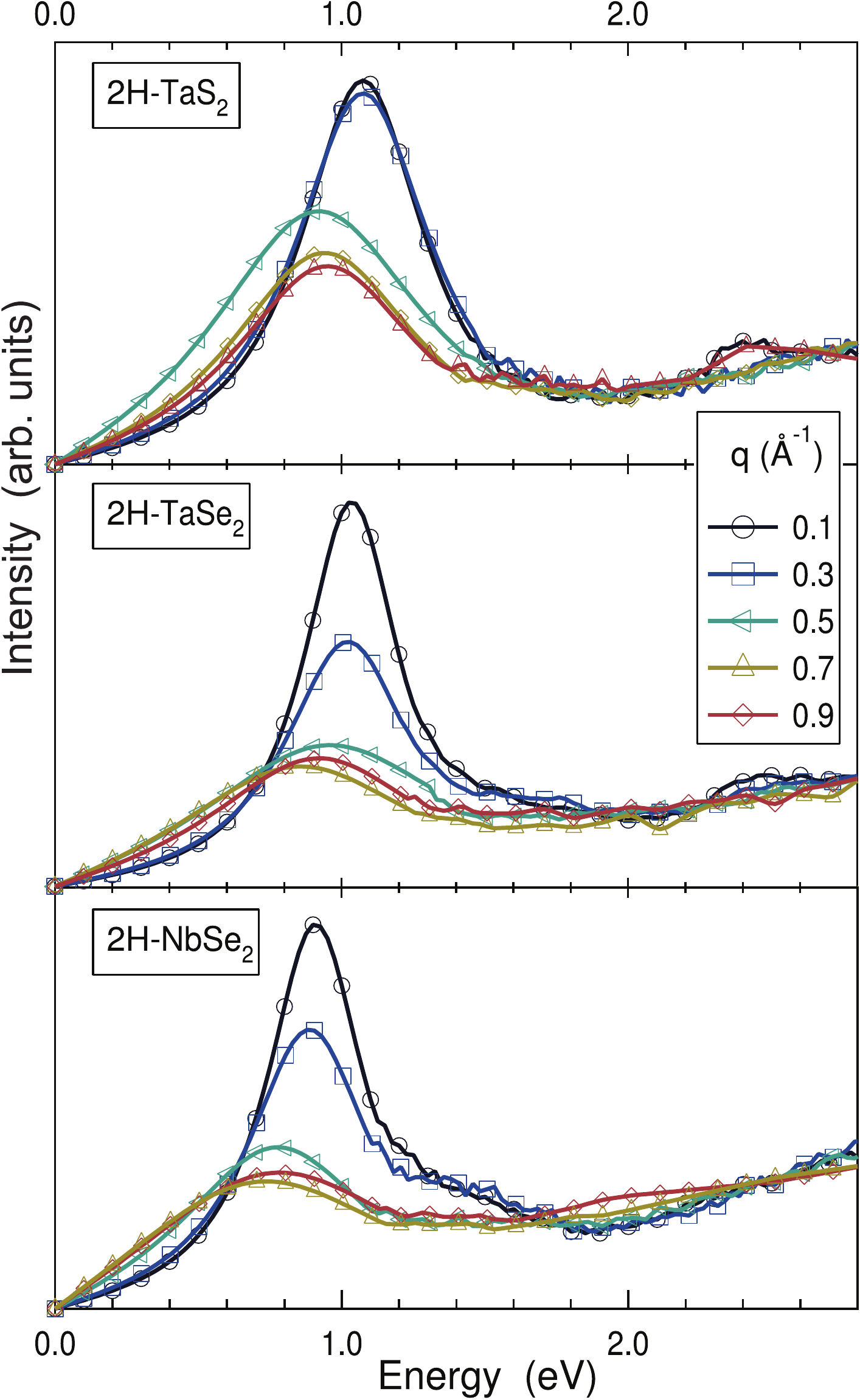}
  \caption{(Color Online) The EELS intensity for the three investigated compounds measured at room temperature. The spectra have been normalized on the high-energy side and the data for \htase\ are partially reproduced from \intextcite{Schuster_Phys.Rev.B_2009}.}
  \label{fig:intensity_overview}
\end{figure}

In \figref{fig:intensity_overview} we show the behavior of the EELS intensity---corrected for the contribution of the quasi-elastic line according to the procedure outlined in \intextcite{Schuster_Phys.Rev.B_2009}---for \htase{}, \htas{} and \hnbse{}, with momentum transfers  along the $\Gamma$K direction of the Brillouin zone (BZ) \cite{Note1}. For small values of the momentum transfer the spectrum is dominated by a pronounced peak around $E=1\unit{eV}$ that corresponds to the plasmon excitation of the conduction electrons. Note that the overall shape of the EELS intensity is consistent with earlier reports \cite{Liang_Philos.Mag._1969,Bell_Adv.Phys._1976}. With increasing momentum transfer the plasmon peak loses strength and becomes successively broadened. In addition, there is a \emph{red}shift of the plasmon peak when moving away from the center of the BZ. 
\begin{figure}[t]
  \centering
  \includegraphics[width=1.0 \columnwidth]{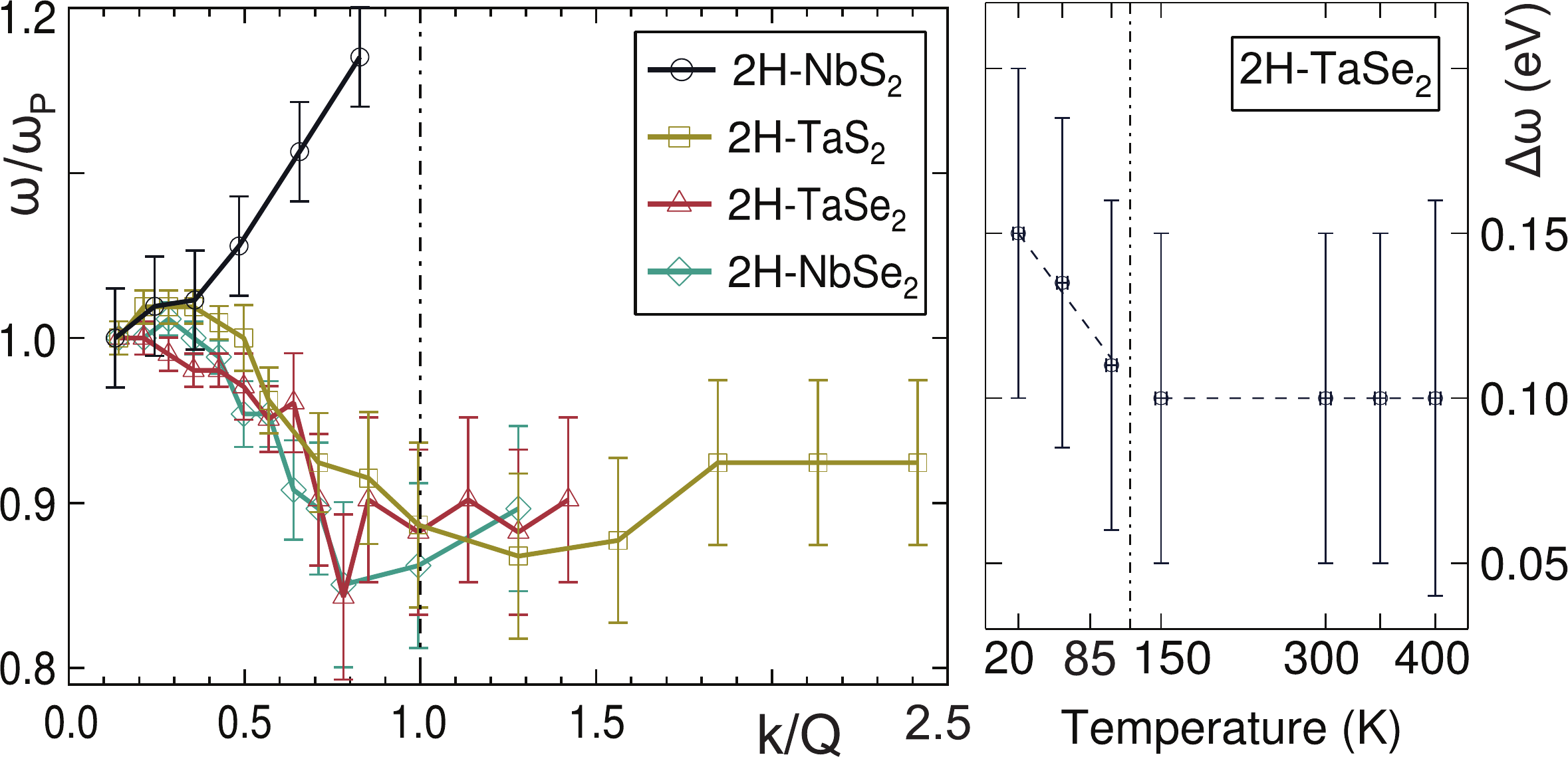}
  \caption{(Color online) Left: The plasmon dispersions for \htas, \htase\ and \hnbse\ as extracted from the peak positions in \figref{fig:intensity_overview} and for \hnbs\ reproduced from \intextcite{Manzke_SolidStateCommun._1981}. The energies have been scaled with respect to the plasma energy for each material, and the momentum transfers with respect to the CDW ordering vector.
Right: Temperature dependence of the plasmon bandwidth for \htase, defined as the difference of the plasmon energy between $q=0.1\rangstrom$ and $q=0.5\rangstrom$. The vertical line indicates the onset of the incommensurate CDW.}
  \label{fig:disp_comparison}
\end{figure}

This is further substantiated in the left panel of \figref{fig:disp_comparison} where we show the plasmon dispersion extracted from the peak positions of the corresponding curves in \figref{fig:intensity_overview}. The behavior for \hnbse, \htase\ and \htas\ is strikingly similar in that the dispersions all have negative slopes for values of the momentum transfer below the CDW ordering vector. In contrast, the dispersion for \hnbs, which does not charge order, is positive everywhere. To further emphasize the link between the negative plasmon dispersion and the presence of CDW order, we show in the right panel of \figref{fig:disp_comparison} the temperature evolution of the plasmon bandwidth (defined here as the difference of the plasmon energy between $q=0.1\rangstrom$ and $q=0.5\rangstrom$). In spite of the rather large error bars due to the increased broadening for higher momentum transfers, it can be seen that the onset of the CDW coincides with an increase in the plasmon bandwidth.

{\it Collective Modes.}--To describe the plasmon excitations in the TMDC, we adopt the formalism developed by Bohm and Pines to separate the plasmonic collective modes from the degrees of freedom of individual electrons in the particle-hole continuum \cite{Pines_Rev.Mod.Phys._1956,Bohm:1953bc}. Starting from the unordered, normal state, the collective part of the electronic Hamiltonian is reduced to:
\begin{align}
H_{\text{plasmon}} = \sum_{k<k_c} \left( \frac{P_k^2}{2 m} + \frac{m}{2} \omp^2(k) X_k^2 \right),
\end{align}
where the coordinates $X_k$ describe the collective motion of all conduction electrons, and the plasmon dispersion is approximated by $\omp^2(k) \simeq \omp^2 + \alpha k^2$, with $\alpha>0$ proportional to the Fermi velocity \cite{Bohm:1953bc}. The collective modes are effectively decoupled from the motions of the individual electrons for wavelengths longer than the electronic screening length $1 / k_c$. 

As a phase transition is approached, collective fluctuations of the order parameter associated with the impending order start to develop. These fluctuations are no longer contained within the electronic screening length, and need to be taken into account in the collective Hamiltonian. This can be done starting from a microscopic theory by introducing an order parameter via a Hubbard-Stratanovich transformation, and tracing over the electronic degrees of freedom. The result is a Ginzburg-Landau energy functional, which may then be included in the semiclassical Hamiltonian \cite{Uchoa:2004ee}: $H=H_{\text{plasmon}} + H_{\text{GL}}$. 

The CDW order parameter $\Psi$ is defined as $\rho = \rho_0 (1 + \Psi)$, and describes fluctuations on top of the average charge density $\rho_0$. In the case of the TMDC, we write $\Psi = \Re \{ \psi_1+\psi_2+\psi_3 \}$ as a sum of three complex variables representing the three components of the observed triple-$q$ CDW. The most general form of the Ginzburg-Landau energy then contains powers of $\alpha$ as well as an interaction term between the CDW components of the form $|\psi_j \psi_{j+1}|^2$, which determine the stability of the triple-$q$ pattern \cite{McMillan:1975wg}. 
The propagation vectors of the CDW components are aligned with the preferred ordering vectors $\vec{Q}_j$ of the charge density wave (as determined from the maximum in the bare susceptibility) by terms proportional to $|\vec{Q}_j \times \vec{\nabla} \psi_j|^2$. If we assume this alignment to be perfect, we can use the three-fold rotational symmetry of the TMDC to write $\psi_j(\vec{r}) = \sum_k \psi(k) e^{i (k/Q) \vec{Q}_j \cdot \vec{r}}$, and the Ginzburg-Landau energy takes on a particularly simple form \cite{McMillan:1975wg, Turgut:e2z5IdWM}:
\begin{align}
F_{\text{GL}} = \sum_k & \left( a + \frac{b}{2} k^2 + \frac{c}{2 k^2} \right) \psi^2(k).
\end{align}
Although the plasmon dispersion is expected to be affected by the presence of CDW fluctuations both above and below $\tcdw$, we will focus on the disordered regime only, so that higher order terms in the expansion may be neglected. It is known that fluctuations of the charge order persist far above the transition temperature in the TMDC \cite{Rossnagel2011,JvW:EPL}, so that $F_{\text{GL}}$ may be assumed to be an adequate approximation of the free energy even beyond the immediate vicinity of $\tcdw$.

Notice that we explicitly included the Coulomb interaction between different parts of the charge density modulation. The contribution from this term is non-negligible when considering the dynamic and collective excitations of the system encountered in the EELS experimental setup. In the present formulation, the balance between the Coulomb interaction and the gradients of $\psi_j(\vec{r})$ also determines the size of the CDW ordering vector $Q=\sqrt[4]{c/b}$. The temperature dependence of the parameter $a$ which drives the CDW transition, is defined as $a = a_0 (T -\tcdw) -\sqrt{b c}$, so that the minimum of $F_{\text{GL}}$ has zero energy at the transition temperature.

{\it Plasmons and Charge Fluctuations.}--Before combining the Ginzburg-Landau description of the CDW order parameter fluctuations with the Hamiltonian for the plasmon modes, it should be realized that both address collective excitations which cause modulations in the total electronic charge distribution. The order parameter $\psi(k)$ can thus not be independent from the collective coordinate $X_k$. In this sense, the interaction between the CDW order parameter and the plasmon dynamics is fundamentally different from that between the plasmon and for example a superconducting order parameter \cite{Turgut:e2z5IdWM}. On the other hand, there is still an important difference between $\psi(k)$ and $X_k$. The plasmon excitation at $E=1\unit{eV}$ involves electrons throughout the conduction band, while the fluctuations of the CDW order parameter occur at the $\unit{meV}$ scale, and concern only a small number of electrons close to $E_F$ \cite{Gruner}. To relate the two quantities to each other, we use the fact that statistical fluctuations of the density in general scale as the square root of the average density of involved particles \cite{Landau_Lifshitz}, to write $n_{\text{CDW}}^2(k) / n_{\text{p}}^2(k)  = n' / n$. The density modulations in the CDW are given by $n_{\text{CDW}}(k)=n' \psi(k)$, with $n'$ the number of electrons involved in the CDW formation per unit volume. The plasmon fluctuations are defined as $n_{\text{p}}(k)=\sqrt{n k^2} X_k$, with $n$ the density of all conduction electrons \cite{Bohm:1953bc}. With this identification of the different collective modes, the semiclassical description of the combined plasmon-CDW system becomes:
\begin{align}
H = \sum_k & \left\{ \frac{P_k^2}{2 m} + \left[\frac{m}{2} \omp^2(k) + \frac{k^2}{2 n'} \left( 2 a + b k^2 \right)\right] X_k^2 \right\}
\end{align}
In this expression we absorbed the Coulomb term of the Ginzburg-Landau theory in the `bare' plasmon energy to avoid double counting of that interaction. Notice also that the kinetic energy of the CDW order parameter, needed to describe the dynamic charge fluctuations, is included in the first term. 
\begin{figure}[t]
  \centering
  \includegraphics[width=0.55 \columnwidth]{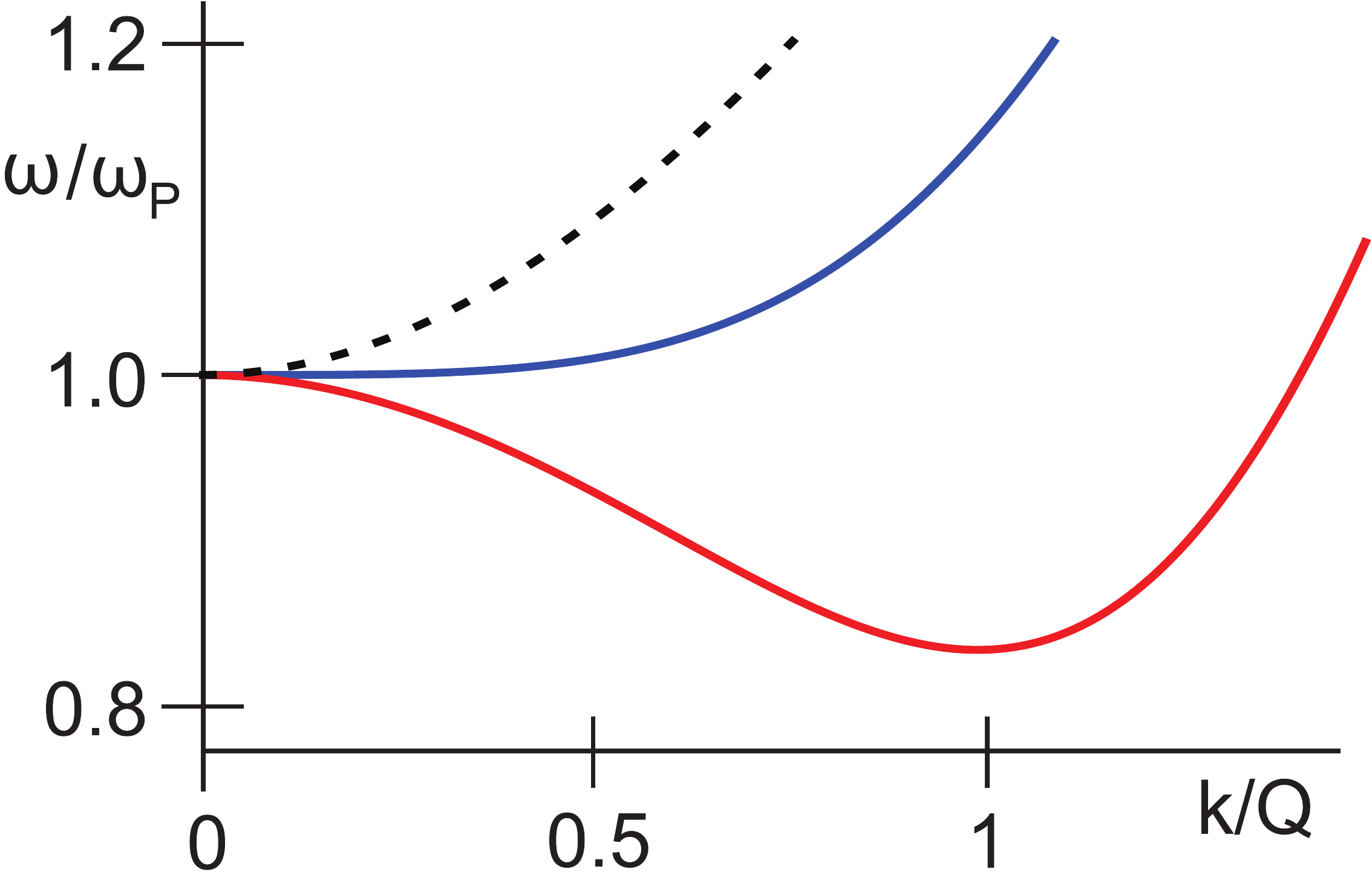}
  \caption{(Color online) The plasmon dispersions of Eq.~\eqref{eq:dispersion}. Depending on the values of the model parameters, the dispersion can be either uniformly positive, of have a negative dispersion at low $k$ followed by a minimum. 
The topmost, dashed line shows the shape of a bare dispersion with $a=b=0$ and $\alpha = 0.75 \omp^2 / Q^2$, for an arbitrary choice of $Q$. The second line is the dispersion for $ a = -\alpha n' m/ 2$, the value at which the second minimum first appears. Here we used $c = b Q^4 = 0.3 n' m \omp^2$. The line on the bottom shows the situation with a well developed minimum for the same value of $c$ and $ a = -\alpha n' m / 2 - \sqrt{b c}$.}
  \label{fig:model_curves}
\end{figure}

From this expression it is immediately clear that the plasmon dispersion close to (but above) the CDW transition temperature is given by:
\begin{align}
\omega(k) = \sqrt{ \omp^2 + \left( \alpha + \frac{2 a}{n' m} \right)k^2 + \left( \frac{b}{n' m} \right)k^4}.
\label{eq:dispersion}
\end{align}
The overall shape of the plasmon dispersion in the vicinity of a CDW instability thus depends on the parameters describing the charge ordering transition. As shown in \figref{fig:model_curves}, the temperature dependence of $a$ implies that at high temperatures, the dispersion is uniformly positive. As the transition towards charge order is approached, a second minimum may develop in addition to the infinite wavelength plasma oscillation which always occurs at $k=0$ with $\omega(0)=\omp$. 
Notice that the second minimum exists only for temperatures $T < \tcdw + (\sqrt{bc}-\alpha) /a_0$. That is, it develops above $\tcdw$ only if the material properties of the CDW system under consideration conspire so that the combination of Coulomb energy and CDW stiffness may overcome the bare plasmon dispersion. This fact may help to explain why a negative dispersion is observed in some CDW materials like the TMDC or for example TTF-TCNQ \cite{Ritsko_Phys.Rev.Lett._1975}, but others like for example the blue bronze, have a strictly positive dispersion \cite{Sing_Phys.Rev.B_1999}.

Using the definition $c=n'^2 e^2 / \epsilon$ for the strength of the Coulomb interaction between different parts of the charge density modulations in the CDW, the plasmon energy at $Q$ can be seen to approach the value $\omega^2 = \omp^2(Q) -n' e^2 / m \epsilon$ at the charge ordering temperature. This result can be straightforwardly interpreted as indicating that the potential energy cost of creating charge modulations with wave vector $Q$ in the electron distribution close to $E_F$ has disappeared at $T=\tcdw$. The kinetic energy associated with the corresponding plasma oscillation and the potential energy cost of displacing electrons deeper in the conduction band are unaffected, and prevent the plasmon mode from softening all the way to zero energy.

{\it Summary and Conclusions.}--In summary, we investigated the dispersion of the charge carrier plasmon in the prototype transition-metal dichalcogenides \htase, \htas\ and \hnbse. For all three compounds we find a negative slope near the center of the Brillouin zone, which contradicts the generic behavior expected for simple metals and also differs significantly from the uniformly positive dispersion in the closely related (but non-CDW) system \hnbs. The occurrence of the negative dispersion in only the TMDC with a CDW transition indicates that an interaction between fluctuations of the charge order and the plasma oscillations is responsible for the deviation from the usual form of the dispersion. A semiclassical description of the plasmon modes in the presence of collective charge fluctuations associated with the vicinity of the CDW transition shows that their combined dynamics indeed gives rise to the observed effects.

Whether the plasmon dispersion in a general CDW system turns fully {\it negative} in the vicinity of the CDW transition depends sensitively on microscopic material properties. Regardless of the sign of the dispersion however, the generality of the theoretical considerations presented here shows that a renormalization of the plasmon dispersion due to collective charge fluctuations will occur in any material on the border of a charge ordering transition. This happens in spite of the very different energy scales characterizing the two phenomena.

{\it Acknowledgements.}--We are grateful to R. Hübel, S. Leger and R. Schönfelder for technical assistance and to J. Fink for making us aware of \intextcite{Manzke_SolidStateCommun._1981}. This project is supported by the DFG project KN393/13, SCHU/2584/1-1, by the Swiss National Foundation for the Scientific Research within the NCCR MaNEP pool, and by the US DOE, Office of Science, under Contract No. DE-AC02-06CH11357.


%

\end{document}